\documentclass[submission,copyright,creativecommons]{eptcs}
\providecommand{\event}{ICE 2020} 
\usepackage{breakurl}             
\usepackage{underscore}           

\usepackage{wrapfig}

\usepackage[square,numbers]{natbib}
\usepackage{stackengine}
\usepackage{mathtools}

\usepackage{float}
\usepackage{color, colortbl}
\usepackage{amsthm}
\usepackage{amsmath,amssymb}
\usepackage{cleveref}
\crefformat{table}{Table}
\usepackage{tocloft}
\usepackage{proof}
\usepackage{theoremref}
\usepackage{enumitem}
\newtheorem{example}{Example}[section]

\title{A type language for message passing component-based systems}

\author{
  Zorica Savanovi\' c
  \institute{IMT School for Advanced Studies, Lucca, Italy}
  \and
Letterio Galletta
  \institute{IMT School for Advanced Studies, Lucca, Italy}
  \and
   Hugo Torres Vieira
  \institute{C4 - University of Beira Interior, Rua Marqu\^es d'\'Avila e Bolama, 6201-001, Covilh\~a, Portugal}
}

\def\titlerunning{A type language for message passing component-based systems}
\def\authorrunning{Z.~Savanovi\' c, L.~Galletta \& H.T.~Vieira}

\cftpagenumbersoff{figure}
\cftpagenumbersoff{table}

\begin{document}
\maketitle


\newcommand{\portal}{$\mathbf{Portal}$}
\newcommand{\storage}{$\mathbf{Storage}$}
\newcommand{\re}{$\mathbf{RE}$}
\newcommand{\girc}{$\mathbf{G_{IRC}}$}
\newcommand{\projection}[2]{#1\downarrow_{#2}}

\newtheorem{proposition}{Proposition}[section]
\newtheorem{convention}{Convention}[section]
 \newtheorem{theorem}{Theorem}[section]
  \newtheorem{lemma}{Lemma}[section]
  \newtheorem{definition}{Definition}[section]
\newtheorem{prop}{Property}[section]
\renewcommand{\figurename}{Figure}
\newcommand{\op}{\textbf{inc}}
\newcommand{\B}{\textbf{B}}
\newcommand{\ren}{\textbf{ren}}
\newcommand{\C}{\textbf{C}}
\newcommand{\D}{\textbf{D}}
\newcommand{\context}{\mathcal{C}}
\newcommand{\transitive}{\overset{*}{<}}
\newcommand{\elem}{\textbf{elements}}
\newcommand{\bc}{\overset{*}{K}}
\newcommand{\of}{\textbf{$F_0$}}
\newcommand{\dc}{F_{dc}}
\newcommand{\tdi}{F_{tdi}}
\newcommand{\td}{F_{td}}
\newcommand{\bnd}{F_{bnd}}
\newcommand{\dpc}{F_{dp}}
\newcommand{\rep}{\mathit{rep}}
\newcommand{\fp}{\mathit{fp}}
\newcommand{\vf}{\mathit{vf}}
\newcommand{\W}{\textbf{W}}
\newcommand{\rel}[2]{\diamond^{#1}_#2}
\newcommand{\pr}{\textbf{pr}}
\newcommand{\A}{\textbf{A}}
\newcommand{\itype}{\mathcal{T}}
\newcommand{\iD}{\mathcal{D}  }
\newcommand{\iC}{\mathcal{C}}
\newcommand{\M}{\mathcal{M}}
\newcommand{\iN}{\mathcal{N}}
\newtheorem*{remark}{Remark}

\newcommand{\inx}{\xrightarrow{\text{$\depen{x?}{b}$}}}
\newcommand{\outy}{\xrightarrow{\text{$\depen{y!}{b}$}}}
\newcommand{\outyone}{\xrightarrow{\text{$\depen{y!}{b_1}$}}}

\newcommand{\io}[2]{\xrightarrow{\text{$\depen{#1}{#2}$}}}

\newcommand{\inxp}{\xrightarrow{\text{$\depen{x?}{b'}$}}}
\newcommand{\inputx}{\xrightarrow{\text{$x?$}}}
\newcommand{\tlp}[2]{#1\!:\!#2}
\renewcommand{\delta}{\sigma}

\newcommand{\N}{\mathbb{N}}
\newcommand{\Z}{\mathbb{Z}}
 \newcommand{\piv}{\Psi}
 \newcommand{\ic}{\Omega}
 \newcommand{\stl}{\textbf{list}}
 \newcommand{\fn}{\textbf{\textit{val}}(x,y,\C,LP,F)}
 \newcommand{\dt}{\D_t(\C,F,LP,y)}
 \newcommand{\dd}{\D_d(\C,F,y)}
  \newcommand{\dtp}{\D_t(\C,F,LP,y')}
 \newcommand{\ddp}{\D_d(\C,F,y')}
 \newcommand{\FN}{\textbf{\textit{val}}}
 \newcommand{\dtr}{\D_t(\C_r,F,LP,\overline{y})}
 \newcommand{\ddr}{\D_d(\C_r,F,\overline{y})}
  \newcommand{\dtrp}{\D_t(\C_r',F,LP,\overline{y})}
 \newcommand{\ddrp}{\D_d(\C_r',F,\overline{y})}

\newcommand{\fig}{\textbf{Figure} \ref}

\newcommand{\partofrecursion}{\[\rep (LP)\overset{\triangle}{=}\{z|LP=\context[\mu \recvar X.\context'[\tlp{z?}{\bt}.LP']]\lor LP=\context[\mu \recvar X.\context'[\tlp{z!}{\bt}.LP']]\}\]}

\newcommand{\freeports}{\[\fp(LP)\overset{\triangle}{=} \{z|LP= \context[\tlp{z?}{\bt}.LP'] \lor LP=\context[\tlp{z!}{\bt}.LP'] \lor LP=\context[\mu \recvar X.\context'[\tlp{z?}{\bt}.LP']] \lor LP=\context[\mu \recvar X.\context'[\tlp{z!}{\bt}.LP']]\}\]}

\newcommand{\FX}{F^{i}}
\newcommand{\FY}{F^{o}}

\newcommand{\pev}[2]{#1\!:\!#2}
\newcommand{\ind}[1]{#1:\Omega}
\newcommand{\T}{T}

\newcommand{\types}[2]{<#1;#2>}
\newcommand{\constr}[4]{\depen{#1}{#2}:#3:[#4]}
\newcommand{\depen}[2]{#1(#2)}
\newcommand{\bt}{b} 
\newcommand{\wbigcup}{\mathop{\bigcup}\displaylimits}
\newcommand{\tcount}[2]{\textit{count}(#1,#2)}

\newcommand{\inpconstset}{T5}
\newcommand{\inpnotdom}{T1}
\newcommand{\inpinit}{T2}

\newcommand{\inppev}{T3}

\newcommand{\outdepset}{T6}
\newcommand{\internal}{T4}

\newcommand{\iinpconstset}{\itype 5}
\newcommand{\iinpnotdom}{\itype 1}
\newcommand{\iinpinit}{\itype 2}

\newcommand{\iinppev}{\itype 3}

\newcommand{\ioutdepset}{\itype 6}
\newcommand{\iinternal}{\itype 4}

\newcommand{\sinpconf}{InpConf}
\newcommand{\soutconf}{OutConf}
\newcommand{\sendconf}{EndConf}
\newcommand{\svarconf}{VarConf}
\newcommand{\srecconf}{RecConf}

\newcommand{\oy}{\overline{y}}



\newcommand{\pp}{\ |\ }
\newcommand{\m}[1]{\mathsf{#1}}

\newcommand{\subst}[2]{\{{}^{#1}\! / \!{}_{#2}\}}
\newcommand{\lto}[1]{\ \mathrel{\stackrel{{\;\;#1\;\;}}{\mbox{\rightarrowfill}}}\ }
\newcommand{\eval}{\downarrow}
\newcommand{\dom}{\m{dom}}


\newcommand{\inl}{{\mathtt{inl}}}
\newcommand{\inr}{{\mathtt{inr}}}

\newcommand{\interface}[2]{[{#1}\,\rangle\,{#2}]}
\newcommand{\chorboxb}[3]{\interface{#1}{#2} \{ #3 \}}
\newcommand{\chorbox}[6]{\interface{#1}{#2} \{ #6; #5; #3; #4\}}

\newcommand{\lbinderp}[2] {   {#1} = {#2}}

\newcommand{\role}[1]{\mathsf{#1}}


\newcommand{\gcom}[3]{\role{#1}\!\!\!\lto{#2}\!\!\!\role{#3}}
\newcommand{\gchoice}[5]{\role{#1}\!\!\!\lto{#2}\!\!\!\role{#3}(#4,#5)}
\newcommand{\recvar}[1]{\mathbf{#1}}
\newcommand{\gend}{\mathbf{end}}

\newcommand{\lab}{\ell}
\newcommand{\gval}[2]{#1,#2}
\newcommand{\labout}[3]{#1!#2\langle#3\rangle}
\newcommand{\labinp}[3]{#1?#2\langle#3\rangle}

\newcommand{\roleport}[2]{\role{#1}.{#2}}
\newcommand{\roleas}[2]{\role{#1}\!=\!#2}


\newcommand{\lfrom}[1]{\mathrel{\stackrel{{\;\;#1\;\;}}{\mbox{\leftarrowfill}}}}
\newcommand{\dbinder}[4] {   \role{#2}.{#3}\lfrom{#1} {#4} }
\newcommand{\lbinder}[3] {   {#1} \,=\, {#3}\,\langle\,{#2} }
\newcommand{\fbinder}[2]{ #1\leftarrow #2 }


\newcommand{\tout}[3]{#1!#2. #3}
\newcommand{\tinp}[3]{#1?#2. #3}
\newcommand{\tchoice}[3]{#1 \oplus (#2, #3)}
\newcommand{\tbranch}[3]{#1\, \& (#2, #3)}
\newcommand{\tend}{\gend}

\newcommand{\chot}{{\mathtt{ChoT}}}


\newcommand{\projc}[4]{#1\! \downarrow_{#2,#3,#4}}
\newcommand{\labenv}{\gamma}

\makeatletter
\newcommand{\xdashrightarrow}[2][]{\ext@arrow 0359\rightarrowfill@@{#1}{#2}}
\newcommand{\xdashleftarrow}[2][]{\ext@arrow 3095\leftarrowfill@@{#1}{#2}}
\newcommand{\xdashleftrightarrow}[2][]{\ext@arrow 3359\leftrightarrowfill@@{#1}{#2}}
\def\rightarrowfill@@{\arrowfill@@\relax\relbar\rightarrow}
\def\leftarrowfill@@{\arrowfill@@\leftarrow\relbar\relax}
\def\leftrightarrowfill@@{\arrowfill@@\leftarrow\relbar\rightarrow}
\def\arrowfill@@#1#2#3#4{%
  $\m@th\thickmuskip0mu\medmuskip\thickmuskip\thinmuskip\thickmuskip
   \relax#4#1
   \xleaders\hbox{$#4#2$}\hfill
   #3$%
}
\makeatother

\begin{abstract}
Component-based development is challenging in a distributed setting, for starters considering programming a task may involve the assembly of loosely-coupled remote components. In order for the task to be fulfilled, the supporting interaction among components should follow a well-defined protocol. In this paper we address a model for message passing component-based systems where components are assembled together with the protocol itself. Components can therefore be independent from the protocol, and reactive to messages in a flexible way. Our contribution is at the level of the type language that allows to capture component behaviour so as to check its compatibility with a protocol. We show the correspondence of component and type behaviours, which entails a progress property for components.

\end{abstract}


\section{Introduction}\label{introduction}

Code reusability is an important principle to support the development of software systems in a cost-effective way. It is a key principle in Component-Based Development (CBD)~\cite{Mcllroy}, 
where the idea 
is to build systems relying on the composition of loosely-coupled and independent units called components.


 The motivations behind CBD are, on the one hand, to increase development efficiency and lower the costs (by building a system from pre-existing components, instead of building from scratch), and on the other hand, to improve quality of the software for instance to what concerns software errors (components can be tested over and over again in different contexts). Consider, for example, microservices (see, e.g.,~\cite{microservices}) that have been recently adopted by massively deployed applications such as Netflix, eBay, Amazon and Uber, and that are reusable 
distributed software units. In such a distributed setting, composing software elements necessarily involves some form of communication scheme, for instance based on message passing.


In order for the functionality to be achieved, communication among components should follow a well-defined protocol of interaction, that may be specified in terms of some choreography language like, for example, WS-CDL~\cite{WS} or the choreography diagrams of BPMN~\cite{BPMN}. 
A component should be able to carry out a certain sequence of I/O actions in order to fulfil its role in the protocol. One way to accomplish this is to implement a component in a way that prescribes a strict sequence of I/O actions, that should precisely match the actions expected by the protocol. However, this choice interferes with reusability, since such a component can be used only in an environment that expects that exact sequence of communication actions. 
For instance, if a component receives an image and outputs its classification just once, what will happen if we need to use this component in a context that requires the classification is sent multiple times? 


In contrast,
a more flexible design choice inspired by reactive programming is to design components so that they can respond to external stimulus without any specific I/O sequence. The reactive programming principle for building such components is to consider that as soon as the data is available, it can be received or emitted. 
For example, we can design a component that is able to output a classification after receiving an image, as long as required. 
In such a way, reusability is promoted since such components can be used in different environments thanks to the flexibility given by the reactive behaviour. However, such a flexibility at the composition level may be too wild if all components are able to send / receive data as soon as it is available. Hence, there is the need to discipline the interactions at the level of the environment where the composition takes place.
What if, for example, we have different images that need to be classified and the classifying component is continuously emitting the result for the first image?  


Carbone, Montesi and Vieira~\cite{Hugo} proposed a language that supports the development of distributed systems by combining the notions of reactive components with choreographic specifications of communication protocols~\cite{fabrizio}. The proposal considers components that can dynamically send / receive data as soon as it is available, while considering that an assembly of components is governed by a protocol. Hence, among all the possible reactions that are supported by the composed components, the only ones that will actually be carried out are the ones allowed by the protocol. A composition of components defines itself a component that can be further composed (under the governance of some protocol) also providing a reactive behaviour. This approach promotes reusability thanks to the flexibility of the reactive behaviour. For instance, by abstracting from the number of supported reactions, e.g., if a component can (always) perform a computation reacting to some data inputs, then it can be used in different protocols that require such computation to be carried out a different number of times; by abstracting from message ordering, e.g., if a component needs some values to perform a computation, it may be used with any protocol that provides them in any order. 

Component implementations should be hidden, so it shouldn't be necessary to inspect the inner workings in order to asses if it is usable in a determined context for the purpose of \textit{off-the-shelf} substitution or reuse. Hence, a component should be characterised with a signature that allows checking its compatibility when used in a composition. In particular, it must be ensured that each component provides (at least) the behaviour prescribed by the protocol in which the component participates. Carbone et al.~\cite{Hugo} propose a verification technique that ensures communication safety and progress. However, the approach requires checking the implementation of components each time the component is put in a different context, i.e., each time that the component is used ``off-the-shelf" we need to check if the reactions supported by the component are enough to implement its part in the protocol.

 In this work we consider a different approach, avoiding the implementation check each time a component is to be used. Firstly, we introduce a type language that characterises the reactive behaviour of components. Secondly, we devise an inference technique that identifies the types of components, based on which we can verify whether the component provides the reactive behaviour required by a context. The motivation is in tune with reusability: once the component's type is identified, there is no further need to check the implementation, because the type is enough to describe ``what the component can do". 
 

Our types specify the ability of components to receive values of a prescribed basic type. Moreover, they track different kinds of dependencies, for instance that certain values to be emitted always (for each output) require a specific set of inputs (dubbed \emph{per each value} dependencies). Our types can also describe the fact that a component needs to be, in some sense, initialised by receiving specific values before proceeding with other reactive behaviour (dubbed \emph{initial} dependencies). Furthermore, our types also identify constraints on the number of values that a component can send.
Finally, we ensure the correctness of our type system by proving that our extraction procedures are sound with respect to the semantics of the Governed Components (GC) language~\cite{Hugo}, considering here the choice-free subset of the GC language and leaving a full account of the language to future work. Moreover, we ensure that whenever a type of a component prescribes an action, a component will not be stuck, i.e., it will eventually carry out the matching action.

The rest of the paper is organised as follows. We first present the GC language in Section~\ref{background}. Then in Section~\ref{example} we intuitively introduce our type language through a motivating example based on AWS Lambda~\cite{aws} where we point out different scenarios that might occur while composing components and how our types allow to describe certain behavioural patterns. Section~\ref{language} introduces the syntax and the semantics of our types. Then, we define the type extraction for base components in Section~\ref{sec:base}, whereas the type extraction for composite components in Section~\ref{sec:composite}. In Section~\ref{safety} we present our results. Section~\ref{finish} concludes, discusses related work and gives some future issues.

%


\section{Background: Governed Components Language}\label{background}

In this section we briefly introduce the GC language, focusing on the main points that allow to grasp the essence of the model and to support a self-contained understanding of this paper. 
We refer the interested reader to~\cite{Hugo} for a full account of the language.
The syntax of the (protocol choice-free fragment of the) GC language is given in Table~\ref{syntaxcomponents}. There are two kinds of components ($K$): base and composite. Both kinds interact with the external environment by means of input and output ports exposed as the component's interface. Besides of the interface, components are defined by their implementation. 

\begin{table}[t]
    \centering
    \begin{tabular}{r l| r l| r l}
    
    \multicolumn{2}{l|}{Components} &  \multicolumn{2}{l|}{Local Binders} &  \multicolumn{2}{l}{Protocol} \\
    
     $K::=$ & $\chorboxb {\tilde x}{\tilde y}{L}$ (base) & $L::=$ & $\lbinderp{y} {\mathit{f}(\tilde{x})}$ & 
     $G ::=$ & $\gcom p \lab{\tilde  q};G$ (communication)\\

     & $\chorbox{\tilde x}{\tilde y}{D}{\role r[F]}{R}{G}$ (composite) &
     & $L,L$ & 
     & $\mu \recvar{X}.G$ (recursion)\\

     & & & & & $\recvar{X}$ (recursion variable)\\

     & & & & & $\gend$ (termination)\\
     
     \hline
      \multicolumn{2}{l|}{Role assignments} &  \multicolumn{2}{l|}{Distribution Binders} &  \multicolumn{2}{l}{Forwarders}\\
     $R::=$ & $\role p=K$ & 
   $D::=$ & $\dbinder{\lab}{p}{x} {\roleport{q}{y}}$ &
     $F::=$ & $\fbinder zw $ \\
     & $R,R$ & & $D,D$ & & $F,F$


    



   
    \end{tabular}
    
    \caption{Syntax of Governed Components}
    
    \label{syntaxcomponents}
    
\end{table}

In the case of a base component the implementation is given by the list of local binders ($\{L\}$). A local binder specifies a function, denoted as $y=f(\tilde{x})$, which is used to compute the output values for port $y$  relying on values received on list of (input) ports $\tilde{x}$. So, we say that component's ability to output a value may depend on the ones received, where instead, components are always able to receive values.
We abstract from the definition of such functions $f$ and assume them to be total. Received values are processed in a FIFO discipline, so queues are added to the local binders at runtime (noted as $\lbinder{y}{\tilde \sigma}{f(\tilde x)}$). Each element ($\sigma$) in a queue ($\tilde \sigma$) is a store defined as a partial mapping from input ports to values ($\tilde \sigma=\sigma_1, \sigma_2, \ldots, \sigma_k$,
where in $\sigma_1$ the oldest values received are stored, in $\sigma_2$ the second-oldest values, and so on and so forth up to $\sigma_k$). E.g., if $y=f(x_1,x_2)<\cdot$ and the component receives $v_1$ and $v_2$ in that order on port $x_1$ and $v_3$, $v_4$ and $5$ on port $x_2$. 
The queue at this point has three mappings $(x_1\rightarrow v_1, x_2 \rightarrow v_3) , (x_1 \rightarrow v_2, x_2 \rightarrow v_4), (x_2\rightarrow 5)$ where two are ``complete'' to compute the function and one is not.

The implementation of a composite component, represented by $\{G;R;D;\role r[F]\}$, is an assembly of subcomponents whose interaction is governed by a protocol ($G$). 
The set of subcomponents are given in $R$ together with their \emph{roles} in the interaction (e.g., $\role r=K$ denotes that component $K$ is assigned to role $\role r$). Composite components also specify a list of (distribution) 
binders ($D$) that provide an association between the messages exchanged in the protocol ($\lab$) and the ports ($x,y$) of the components (e.g., $\dbinder{\lab}{p}{x}{\roleport{q}{y}}$ states that a message with a label $\lab$ is emitted on port $y$ of the component assigned to role $\role q$, and to be received on port $x$ of the component assigned to role $\role p$). Ports are uniquely associated to message labels ($\lab$) in such way that each communication step in the protocol has a precise mapping to a port, i.e., all values emitted on a port will be carried in messages with the same label and all values received on a port will be delivered in messages with the same label. E.g., every time a value is emitted from some port $y$ it will be carried in a message labelled with e.g., $\lab$ and a value delivered on some port $x$ is delivered on a message with the same label $\lab$. Some other label cannot be attached to $y$, otherwise the association would not be unique.
The $class$ message label (or some other) cannot be attached to $y_p$ otherwise the association would not be unique.  Finally, subterm $\role r[F]$ is used to specify the externally-observable behaviour: the (only interfacing) subcomponent responsible for the interaction with the external environment is identified (by its role $\role r$) and the respective connection between ports is provided by forwarders ($F$). The idea is that values received on the (input) ports of the composite component are directly forwarded to the (input) ports of the interfacing subcomponent, and values emitted on the (output) ports of the interfacing subcomponent are forwarded to the (output) ports of the composite component. For example, the term $\fbinder {x'}x$ is for forwarding an input, and the term $\fbinder y{y'}$ is for forwarding an output ($x$ and $y$ are the ports of the composite component).  


Protocol specifications prescribe the interaction among a set of parties identified by roles. Communication term $\gcom p \lab{\tilde  q};G $ specifies that role $\role p$ sends the message labelled  $\lab$  to the (nonempty) set of roles $\tilde{\role q}$, after which the protocol continues as specified by $G$. The difference between this work and~\cite{Hugo} is the absence of branching. Then we have terms $\mu \recvar{X}.G$ and $\recvar X$ for specifying recursive protocols. Finally, term $\gend$ defines the termination of the protocol.   

We now present the operational semantics of the GC in terms of a labelled transition system (LTS).  We denote by $K \lto{\lambda} K'$ that a component $K$ evolves in one computational step to $K'$, where observations are captured by labels defined as follows $\lambda::=x?v\ |\ y!v\ |\ \tau$. Transition label $x?v$ represents an input on port $x$ of a value $v$, label $y!v$ captures an output on port $y$ of a value $v$, and label $\tau$ denotes an internal move.

The rules that describe the behaviour of components are presented in two parts, addressing base and composite components separately. We present only the main rules, the full semantics can be found in~\cite{Hugo}.   

\begin{table}[h!]
\centering
\begin{tabular} {c c c}

        \infer[\m{OutBase}]
        {
        \chorboxb {\tilde x}{\tilde y}{L}
        \lto{y!{v}{}}
        \chorboxb {\tilde x}{\tilde y}{L'}
        }
        {
        L
        \lto{y!{v}{}{}}
        L'
        & y \in \tilde y
        }
 & &

        \infer[\m{InpBase}]
        {
        \chorboxb {\tilde x}{\tilde y}{L}
        \lto{x?v}
        \chorboxb {\tilde x}{\tilde y}{L'}
        }
        { L \lto{x?v} L' & x \in \tilde x }
\\

\end{tabular}
  \caption{Semantics of base components}
  \label{tab:gcsemanticsbase}
\end{table}

Rules  $\m{OutBase}$ and $\m{InpBase}$ that are given in Table~\ref{tab:gcsemanticsbase} capture base component behaviour, and are defined relying on transitions exhibited by local binders, denoted $L \lto{\lambda} L'$.
Rule $\m{OutBase}$ states that if
local binders $L$  can exhibit an output of value $v$ on port $y$, where $y$ is part of the component's interface,
then the corresponding output can be exhibited by the base component.
Rule $\m{InpBase}$ follows the same lines.

\begin{table}
\centering
\begin{tabular}{c}

	\infer[\m{OutComp}]
	{
	\chorbox {\tilde x}{\tilde y}{ D}{\role{r}[ F]}{\roleas{r}{K},  R}{G}
	\lto{{y}!{v}{}}
	\chorbox {\tilde x}{\tilde y}{ D}{\role{r}[ F]}{\roleas{r}{K'},  R}{G}
	}	
	{
	K\lto{{z}!{v}{}{}} K'
	\qquad
	F = \fbinder {y}z ,  F'
	\qquad
	y \in \tilde y
	}\\

        \infer[\m{InpComp}]
        {\chorbox{\tilde x}{\tilde y}{ D}{\role{r}[ F]}{\roleas{r}{K}, R}{G}
	\lto{{x}?v}
        \chorbox{\tilde x}{\tilde y}{ D}{\role{r}[ F]}{ \roleas{r}{K'}, R}{G}}        
	{ 
	K \lto{{z}?v} K'  \qquad  F = \fbinder z{x},  F' \qquad x \in \tilde x
	}\\

	\infer[\m{Internal}]
	{\chorbox{\tilde x}{\tilde y}{ D}{\role{r}[ F]}{\roleas{s}{K},  R}{ G}
	\lto{\tau}
	\chorbox{\tilde x}{\tilde y}{ D}{\role{r}[ F]}{ \roleas{s}{K'},  R}{G}}	
	{K \lto{\tau} K'}
\\

  \infer[\m{OutChor}]
        {\chorbox{\tilde x}{\tilde y}{ D}{\role{r}[ F]}{\roleas{p}{K}, R}{G}
	\lto{\tau}
        \chorbox{\tilde x}{\tilde y}{ D}{\role{r}[ F]}{ \roleas{p}{K'}, R}{G'}}        
	{K \lto{{u}!{v}{}} K' \qquad 
	D = \dbinder{\lab}{q}{z} {\roleport{p}{u}}, D'
        \qquad
        G \lto{\labout{\role p}{\lab}{v}} G'
        }
\\

        \infer[\m{InpChor}]
        {\chorbox{\tilde x}{\tilde y}{ D}{\role{r}[ F]}{ \roleas{q}{K},  R}{G}
	\lto{\tau}
        \chorbox{\tilde x}{\tilde y}{ D}{\role{r}[ F]}{\roleas{q}{K'},  R}{G'}}        
	{
        K \lto{{z}?v} K' \qquad
         D = \dbinder{\lab}{q}{z} {\roleport{p}{u}},  D' \qquad
        G \lto{\labinp{\role q}{\lab}{v}} G'
	}

\end{tabular}
  \caption{Semantics of composite components}
  \label{tab:gcsemanticscomposite}
  
\end{table}

Notice that the transition of the local binder specifies a final configuration $L'$ which is accounted for in the evolution of the base component. We omit here the rules for local binders (see~\cite{Hugo}) and provide only an informal account for them. Essentially, a (runtime) local binder $\lbinder{y}{\tilde \sigma}{f(\tilde x)}$ is always receptive to an input $x?v$: if $x$ is not used in the function ($x \not \in \tilde x$) then value $v$ is simply discarded; otherwise, the value is added to the (oldest) entry in mapping queue $\tilde \sigma$ that does not have an entry for $x$ (possibly originating a new mapping at the tail of $\tilde \sigma$). All local binders in $L$ synchronise on an input, so each local binder will store (or discard) its own copy of the value. Instead, local binder outputs are not synchronised among them: if a local binder outputs a value, other local binders will not react. For this to happen, the oldest mapping in queue $\tilde \sigma$ must be complete (i.e., assign values to all of $\tilde x$) at which point function $f$ may be computed, the result which is then carried in the transition label (i.e., the $v$ in $y!v$).

We now introduce the rules that capture the behaviour of the composite components, displayed in Table~\ref{tab:gcsemanticscomposite}. 
Notice that a composite component may itself be used as a subcomponent of another composition (of a ``bigger'' component), and base components provide the syntactic leaves.
%
Rules $\m{OutComp}$ and $\m{InpComp}$ capture the interaction of a composite component with an external environment, realised by the interfacing subcomponent. The role assignment $\roleas{r}{K}$ captures the relation between component $K$ and role $r$, which is specified as the interfacing role ($\role{r}[ F]$). Rule $\m{OutComp}$ allows for the interfacing component $K$ to send a value $v$ to the external environment via one of the ports of the composite component ($y$). Notice that the connection between the port of the interfacing component ($z$) and the port of the composite component ($y$) is specified in a forwarder ($F = \fbinder {y}z ,  F'$). Rule $\m{InpComp}$ follows the same lines to model an externally-observable input.
Rule $\m{Internal}$ allows for internal actions in a subcomponent ($K$), where the final configuration ($K'$) is registered in the final configuration of the composite component.

Rules $\m{OutChor}$ and $\m{InpChor}$ capture the interaction among subcomponents of a composite component. 
Rule $\m{OutChor}$ addresses the case when a component is sending a message to another one. The premises, together with role assignment $\roleas{p}{K}$, establish the connection among sender component $K$, the component port $u$, sender role $\role{p}$, and  message label $\lab$.
Premise $K \lto{{u}!{v}{}} K'$ says that the sender component ($K$) can perform an output of value $v$ on port $u$. Premise $D = \dbinder{\lab}{q}{z} {\roleport{p}{u}}, D'$ says that the distribution binders specify the (unique) relation between port $u$ of sender role $\role{p}$ and message label $\lab$ (receiver role $\role{q}$ and associated port $z$ are not important here).
The last premise $G \lto{\labout{\role p}{\lab}{v}} G'$ realises the component governing the protocol, i.e., saying that the communication is only possible if the protocol prescribes it. Namely, the premise says that the protocol exhibits an output of a value $v$ carried in message $\lab$ from role $\role{p}$. 
The rules for protocol transitions are presented in~\cite{Hugo}. Naturally which semantics has an impact on our technical development (namely regarding end-point projection in local protocols), but to some extent can be addressed in a modular way (i.e., up to the existence of the end-point projection). 
Notice that the transitions of component $K$ and  protocol $G$ specify final configurations $K'$ and $G'$  which are accounted for in the evolution of the composite component.
 
 
 Rule $\m{InpChor}$ is similar, but instead of message sending, it addresses the case when a subcomponent receives a message from another subcomponent. The premises are equivalent to the ones for Rule $\m{OutChor}$, but now regard reception. Namely, by saying that receiving component $K$ can exhibit the respective input transition, that the distribution binder specifies the relation of message label $\lab$ with receiver role $\role{q}$ and port $z$, and that protocol $G$ prescribes the input of a value.

     
    





\section{Motivating example: microservices for Image Recognition System}\label{example}

In order to further motivate GC and also to introduce our typing approach, we now informally discuss an example inspired by a microservices scenario~\cite{aws} that addresses an Image Recognition System ($\role {IRS}$). The basic idea is that users upload images and receive back the resulting classification. Moreover, users can get the current running version of the system whenever desired. 
The $\role {IRS}$ is made of two microservices, $\role {Portal}$ and $\role {Recognition}$ $\role {Engine}$ ($\role {RE}$), that interact according to a predefined protocol.

The task is achieved according to the following workflow: $\role {Portal}$ sends the $\mathit{image}$ loaded by a user to  $\role {RE}$ to be processed. When $\role {RE}$ service finishes its $\mathit{classification}$, it sends  the $\mathit{class}$ as the result of the $\mathit{classification}$ to $\role {Portal}$. We model the scenario in the GC calculus by assigning to each microservice the corresponding role and using components to represent them. We assign  role $\role {Portal}$ to component $K_{Portal}$ and role $\role {RE}$ to component $K_{RE}$, where $K_{Portal}$ and $K_{RE}$ are base components. 
Interaction between these two components is governed by global protocol $G$, that can be described as: $\gcom {Portal} {image} {RE};\gcom {RE} {class} {Portal}$. This (the part of $G$) protocol exactly specifies the workflow described above:  $\role{Portal}$ sends an $image$ to $\role {RE}$ ($\gcom {Portal} {image} {RE}$) that answers with the computed $class$ ($\gcom {RE} {class} {Portal}$). If we add the termination ($\gend$) we obtain (complete $G$) protocol  $\gcom {Portal} {image} {RE};\gcom {RE} {class} {Portal};\gend$,  which may be described as a one-shot protocol, since the interaction is over ($\gend$) after the components exchange the two messages.

We obtain composite component $K_{IRS}$ by assembling $K_{Portal}$ and $K_{RE}$ together with protocol $G$ that governs the interaction. 
%
%
Below, we show how it is possible to graphically represent component $K_{IRS}$, where we represent  $K_{Portal}$ and $K_{RE}$ as its subcomponents:

   \begin{center}
 \includegraphics[width=0.8\linewidth]{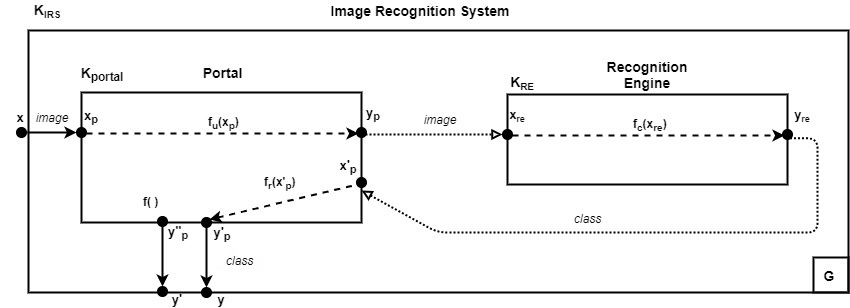} 
\end{center}
  
The subcomponent $K_{Portal}$ is the interfacing component (hence is the only one connected to the external environment via forwarders). We can specify $K_{Portal}$ in the GC language as \vspace{-0.2cm} \[ 
\interface{x_p,x_p'}{y_p,y_p',y_p''}\{y_p=f_u(x_p)<\tilde{\sigma}^{y_p},y_p'=f_r(x_p')<\tilde{\sigma}^{y_p'},y_p''=f()<\cdot\}\]

As previewed in the graphical illustration, from the specification we can see that $K_{Portal}$ component has two input ports ($x_p,x_p'$), three output ports ($y_p,y_p',y_p''$), and three local binders that at runtime are equipped with queues ($\tilde{\sigma}^{y_p}$, $\tilde{\sigma}^{y_p'}$ and empty queue $\cdot$ given that the respective binder does not use any input ports). Notice that the queues are only required at runtime and are initially empty.

The idea of our type description is to provide an abstract characterisation of component's behaviour. Types provide information about the set of input ports, namely the types of values that can be received on them, and about the output ports, namely regarding their behavioural capabilities. In particular, for each output port there are constraints which comprise three pieces of information: what type of values are emitted; what is the maximum number of values that can be emitted; and what are the dependencies on input ports, possibly including the number of currently available values that satisfy the dependency at runtime. 

Informally, the type of $K_{Portal}$ announces the following: In the two input ports $x_p$ and $x_p'$ the component can receive an $\mathit{image}$ and a $\mathit{class}$, respectively (\{$\depen{x_p}{\mathit{image}},\depen{x_p'}{\mathit{class}}\}$).  
Also, 
the type says in $y_p$ the component can emit $\mathit{image}$s and it can do so an unbounded number of times (denoted by $\infty$) as the underlying local binder imposes no boundary constraints. In particular, the local binder can send an $\mathit{image}$ as soon as one is received in $x_p$. Hence, we have a \emph{per each} value dependency of $y_p$ on $x_p$. Formally, we write this constraint as $\constr{y_p}{image}{\infty}{\{\pev{x_p}{N_p}\}}$, where $N_p$ is the number of values received on $x_p$ that are available to be used to produce the output on $y_p$. We may describe constraint $\constr{y_p'}{\mathit{class}}{\infty}{\{\pev{x_p'}{N_p'}\}}$ in a similar way. In constraint $\constr{y_p''}{\mathit{version}}{\infty}{\emptyset}$ there are no dependencies from input ports specified, hence the reading is only that a $\mathit{version}$ can be emitted an unbounded number of times. We may specify the type of $K_{Portal}$ as: \[T_{\mathit{Portal}}=<\{\depen{x_p}{image},\depen{x_p'}{class}\};\{\constr{y_p}{image}{\infty}{\{\pev{x_p}{N_p}\}},\constr{y_p'}{class}{\infty}{\{\pev{x_p'}{N_p'}\}},\constr{y_p''}{version}{\infty}{\emptyset}\}>\]

Composite component $K_{IRS}$ is an assembly of two base components $K_{\mathit{Portal}}$ and $K_{RE}$ whose communication is governed by global protocol $G$. The description of $K_{IRS}$ in GC language is the following: \[K_{IRS}=\chorbox {x}{ y,y'}{D}{\role{Portal}[F]}{\role{Portal}=K_{Portal}, \role {RE}=K_{RE}}{G}\]
where $G$ is the already described one-shot protocol  \[G=\gcom {Portal} {image}{RE}; \gcom {RE} {class}{Portal};\gend\]
Interfacing component $K_{\mathit{Portal}}$ forwards the values from/to the external environment as specified in the forwarders ($F= \fbinder {x_p}{x},  \fbinder {y}{y_p'},  \fbinder {y'}{y''_p}$). 
The forwarding implies that the characterisation of ports $x$, $y$ and $y'$ in the type of $K_{IRS}$ relies on one of the ports $x_p$, $y_p'$ and $y_p''$, respectively, in the type of $K_{\mathit{Portal}}$. 

The type of $K_{IRS}$ then says that it can always input on $x$ values of type $\mathit{image}$ accordingly to the input receptiveness principle.
The constraint for $y'$ is the same as for $y_p''$ since $y_p''$ does not depend on the protocol (in fact it has no dependencies).
However, this is not the case for $y$: in order for a $\mathit{class}$ of an image to be forwarded from $y_p'$ there is a dependency (identified in $T_{\mathit{Portal}}$) on port $x_p'$. 
Furthermore, component $K_{\mathit{Portal}}$ will only receive a value on $x_p'$ accordingly to the protocol specification, in particular upon the second message exchange. Hence, there is also a protocol dependency since the first message exchange has to happen first, so there is a transitive dependency to an $\mathit{image}$ being sent in the first message exchange, emitted from port $y_p$ of component $K_{\mathit{Portal}}$. Finally, notice that $y_p$ depends on $x_p$ which is linked by forwarding to port $x$ of component $K_{IRS}$, thus we have a sequence of dependencies that link $y$ to $x$.

Since we have a one-shot protocol, the communications happens only once, which implies that one $\mathit{class}$ is produced for the first $\mathit{image}$ received. We therefore consider that the dependency of $y$ on $x$ is \textit{initial} (since one value suffices to break the one-shot dependency), and that the maximum number of values that can be emitted on $y$ is 1. This constraint is formally written as $\constr{y}{class}{1}{\{\ind{x}\}}$. The constraint for $y'$ is $\constr{y'}{version}{\infty}{\emptyset}$, where the set of dependencies is empty, i.e., it does not depend on any input. We then have the following type for component $K_{IRS}$: \[T_{IRS}=<\{\depen{x}{image}\};\{\constr{y}{class}{1}{\{\ind{x}\}},\constr{y'}{version}{\infty}{\emptyset}\}>\]

 Let us now assume a recursive version of protocol \[G'=\mu\recvar X.\gcom {Portal} {image}{RE}; \gcom {RE} {class}{Portal};\recvar X\] 
is used instead (i.e., $K_{IRS}'=\chorbox {x}{ y,y'}{D}{\role{Portal}[F]}{\role{Portal}=K_{Portal}, \role {RE}=K_{RE}}{G'}$).
The idea now is that for each $\mathit{image}$ received a $\mathit{class}$ is produced. So, $\mathit{class}$ may be emitted on $y$ an unbounded number of times and the dependency of $y$ on $x$ is of a \emph{per each} kind. Notice that the chain of dependencies can be described as before, but the one-shot dependency from before is now renewed at each protocol iteration.

The constraint for $y$ in this settings is $\constr{y}{class}{\infty}{\{\pev{x}{N_i}\}}$, where $N_i$ captures the number of values received on $x$ that are currently available to produce the outputs on $y$. The constraint for $y'$ is the same as in the previous case. We then have that the type of $K_{IRS}'$ is \[<\{\depen{x}{image}\};\{\constr{y}{class}{\infty}{\{\pev{x}{N_i}\}},\constr{y'}{version}{\infty}{\emptyset}\}>\]
 
Imagine that $K'_{Portal}$ is now a composite component that has an initialisation phase such that, first it receives a message about what kind of classification is required (e.g., ``classify the image by the number of faces found on it''), then it sends it to $K'_{RE}$, after which the uploading and classification of the images can start (all other characteristics remain). Let $x_1$ be the port of $K'_{IRS}$ on which this message is received. 
Let us consider the following protocol \[G''=\gcom {Portal} {kind}{RE}; \mu\recvar X.\gcom {Portal} {image}{RE}; \gcom {RE} {class}{Portal};\recvar X\] where after component $K'_{Portal}$ sends the required kind of  classifications (labelled as $\mathit{kind}$), the communication between $K'_{Portal}$ and $K'_{RE}$ is governed by a recursive protocol as described in the previous example. The type of the component $K'_{IRS}$ is similar to the type from the previous example, but now announces that the output on $y$ requires an initial value to be received on port $x_1$, as the image classification process can only start after that. We then have the type of $K'_{IRS}$ \[<\{\depen{x}{image}\};\{\constr{y}{class}{\infty}{\{\pev{x}{N_i},\ind{x_1}\}},\constr{y'}{version}{\infty}{\emptyset}\}>\]

 
 


\section{A type language for the components}\label{language}

 In this section we define the type language that captures the behaviour of components in an abstract way,
starting with the presentation of the syntax which is followed by the operational semantics. Then we present two procedures that define how to extract the type of a component. The first procedure is for base, and the second one is for composite components. 
\vspace{-0.4cm}

\paragraph{Syntax}

The syntax of types is presented in Table~\ref{tab:typesyntax} and some
explanations follow.
A type $\T$ consists of two elements: a (possibly empty) set of input ports, 
where each one is associated with a basic type $\bt$ (i.e., int, string, etc.), and a (possibly
empty) set of
constraints $\C$, one for each output port. 
Basic types (ranged over by $\bt, \bt_1,\bt_2,\bt^x,\bt^y,\bt', \ldots$) specify the type of the values that can be communicated through ports, so as to ensure that no unexpected values arise. 
Each constraint in $\C$ contains a triple of the form
$\constr{y}{\bt}{\B}{\D}$, which describes the type ($\bt$) of values sent via $y$, the
capability ($\B$) of $y$ and the dependencies ($\D$) of $y$ on the input ports. The set of constraints $\C$ is ranged over $\C_1,\C_2, \dots, \C', \C'', \C^y, \dots$ (likewise other syntactic categories like $N$, $\B$, $\D, \dots$).
Capability $\B$ identifies the upper bound on the number of values that can be sent from
the output port: a natural number $N$ denotes a bounded capability, whereas $\infty$
 an unbounded one. Dependencies are of two kinds:  \textit{per each value} dependencies are of the form $\pev{x}{N}$ and \textit{initial} dependencies are given by $\ind{x}$.
 A dependency $\pev{x}{N}$ says that each value emitted on $y$ requires the reception of one value on $x$, and furthermore $N$ provides the (runtime) number of values available on $x$ (hence, initially $N = 0$).
 Instead, a dependency $\ind{x}$ says that $y$ initially depends on a (single) value
 received on $x$, hence the dependency is dropped after the first input on $x$. 

Note that there are only two kinds of dependencies: a per each value dependency and an initial one. Since we aim at static typing, the dependencies that  appear after the extraction of a type are either $\pev{x}{0}$ or $\ind{x}$, but for the sake of showing our results, we need to capture these values in the evolution of the types. So, we need to capture the number of values available on the input ports, hence we have dependencies of the kind $\pev{x}{1},\pev{x}{2},\dots$ (for $N=1,N=2\dots$).

\begin{remark}
In this work we investigate the mathematical model for the purpose of showing our results, but we may already point towards practical applications.
In particular, for the purpose of the (static) type verification we are aiming at, the \emph{counting} required for the theoretical model would not be involved 
and the component type information available for developers would be as follows:

\begin{itemize}
    \item set of input ports with their basic types
    \item set of constraints for each output port with the following information
    \begin{enumerate}
        \item the basic type associated to the output port
        \item one of two possibilities for output port capability: bounded or unbounded
        \item one of two possibilities for each kind of dependency: per each value or initial
    \end{enumerate}
\end{itemize}

Hence, for the sake of static type checking and from a developers perspective, apart the expected information regarding basic (value) types, the type information would be \verb+y:bounded+ or \verb+y:unbounded+ to what concerns output port capabilities and \verb+x:pereach+ or \verb+x:initial+ to what concerns dependencies.

\end{remark}

\vspace{-0.5cm}
\begin{table}[t]
\centering
\small{
\begin{tabular}{c |c  |c }

Types and input interfaces &   Dependency kinds &  Boundaries\\

$\T\overset{\Delta}{=}\types{X_b}{\C} $
\quad
$X_b\overset{\Delta}{=}\{\depen{x_1}{b_1}, \dots , \depen{x_k}{b_k}\} $

&  $M::=N\ |\ \ic$ 

&
$\B::=N\ |\ \infty$

\\[0.1cm]
\hline

Constraints & Dependencies & \\
 $\C\overset{\Delta}{=}\{\constr{y_1}{\bt_1}{\B_1}{\D_1}, \dots , \constr{y_k}{\bt_k}{\B_k}{\D_k}\}$  
  & $\D\overset{\Delta}{=}\{\pev{x_1}{M_1}, \dots , \pev{x_k}{M_k}\}$ 
  & $k\geq 0 \quad N\in \mathbb{N}_0$

\end{tabular}}

 \caption{Type Syntax}
    \label{tab:typesyntax}
\end{table}

\paragraph{Semantics}

We now define the operational semantics of the type language,
that is required to show that types faithfully capture component behaviour. The semantics is given by the LTS shown in
Table~\ref{tab:semantics}. 
There are four kinds of labels $\lambda$ described by the following grammar: $\lambda::= x?\ |\ \depen{x?}{\bt}\ |\ \depen{y!}{\bt}\  |\ \tau$. Label $x?$ denotes an input on $x$; whereas, label $\depen{x?}{\bt}$ denotes an
input of a value of type $\bt$;
then, label $\depen{y!}{\bt}$ represents an output of a value of type $\bt$;
finally, $\tau$ captures an internal step.

We briefly describe the rules shown in Table~\ref{tab:semantics}.
Rules [\inpnotdom,\inpinit,\inppev] describe inputs of a (single) constraint, while [\internal, \inpconstset, \outdepset] capture type behaviour.
Rule [\inpnotdom] says a constraint for $y$ can receive (and discard) an input on $x$ in case $y$ does not depend on $x$, i.e., if $x$ is not in the domain of $\D$
($dom(\D)=\{x\ |\ \D=\{\pev{x}{M}\}\uplus\D'\}$),  leaving the constraint unchanged.
Rule [\inpinit] addresses the case of an initial dependency, where after receiving the value on $x$ the dependency is removed.
Rule [\inppev] captures the case of a per each value dependency, where after the reception the number of values available on $x$ for $y$ is incremented. 

With respect to type behaviour, 
Rule [\internal] says that the type can exhibit an internal step and remain unchanged, used to mimic component internal steps (which have no impact on the interface).
Rule [\inpconstset] states that if all type constraints can exhibit an input on $x$ and $x$ is part of the type input interface, then the type can exhibit the input on $x$ considering the respective basic type.
Notice that rules [\internal, \inpconstset, \outdepset] say that constraints can always exhibit an input (simply the effect may be different).
Finally, Rule [\outdepset] says that if one of the constraints has all of the dependencies met, i.e., has at least one value for each $x$ for which there is a dependency, and also that the boundary has not been reached (i.e., it is greater than zero), then the type can exhibit the corresponding output implying the decrement of the boundary and of the number of values available in dependencies.
Notice that in order for a port to output a value, there can be no initial dependencies present (which are dropped once satisfied), only per each value dependencies.

\begin{table}[t]

\begin{center}

\small
$\begin{tabular}{c c}

  \infer[{[\inpnotdom]}]{\constr{y}{\bt}{\B}{\D}\xrightarrow{\text{$x?$}}\constr{y}{\bt}{\B}{\D}}{x\notin dom[\D]} &   \infer[{[\inpinit]}]{\constr{y}{\bt}{\B}{\{\ind{x}\}\uplus\D}\xrightarrow{\text{$x?$}}\constr{y}{\bt}{\B}{\D}}{}\\[0.2cm]

\infer[[{\inppev]}]{\constr{y}{\bt}{\B}{\{\pev{x}{N}\}\uplus\D}\xrightarrow{\text{$x?$}}\constr{y}{\bt}{\B}{\{\pev{x}{N+1}\}\uplus\D}}{} & \infer[{[\internal]}]{T\xrightarrow{\text{$\tau$}}T}{}\\[0.2cm]

  \multicolumn{2}{c}{\infer [{[\inpconstset]}]{<\{\depen{x}{\bt^x}\uplus X_b\};\{\constr{y_i}{b_i}{\B_i}{\D_i} | i \in 1,\dots ,k \}>\xrightarrow{\text{$\depen{x?}{\bt^x}$}}<\{\depen{x}{\bt^x}\uplus X_b\};\{\constr{y_i}{b_i}{\B_i}{\D'_i} | i \in 1,\dots ,k \}>}{\forall i\in 1,2,\dots, k &  \constr{y_i}{b_i}{\B_i}{\D_i}\xrightarrow{\text{x?}}\constr{y_i}{b_i}{\B_i}{\D_i'}}}\\[0.2cm]
 \multicolumn{2}{c}{  \infer [{[\outdepset]}]{<X_b;\{\constr{y}{\bt^y}{\B}{ \{\pev{x_i}{N_i}  | i \in 1,\dots ,k \}}\}\uplus \C>\xrightarrow{\text{$\depen{y!}{\bt^y}$}}<X_b;\{\constr{y}{\bt^y}{\B-1}{\{\pev{x_i}{N_i-1}  | i \in 1, \dots ,k \}}\}\uplus \C>}{\forall i\in 1,2,\dots, k& N_i\geq 1  &   \B>0 }}  \\

 \end{tabular}$

\end{center}
\caption {Type Semantics} \label{tab:semantics}
\vspace{-0.4cm}
\end{table}

In the following example and in the rest of the paper (where appropriate) we adopt the following notation:
 $i$ abbreviates the $\mathit{image}$ type, $c$ abbreviates the $\mathit{class}$ type and $v$ abbreviates the $\mathit{version}$ type. 

\begin{example}
We revisit the type of component $K_{\mathit{Portal}}$ shown in Section~\ref{example}
 \[<\{\depen{x_p}{i},\depen{x_p'}{c}\};\{\constr{y_p}{i}{\infty}{\{\pev{x_p}{N_1}\}},\constr{y_p'}{c}{\infty}{\{\pev{x_p'}{N_2}\}},\constr{y_p''}{v}{\infty}{\emptyset}\}>\] for some $N_1$ and $N_2$.
%
Recall also type $<\{\depen{x}{\mathit{i}}\};\{\constr{y}{\mathit{c}}{1}{\{\ind{x}\}},\constr{y'}{\mathit{v}}{1}{\emptyset}\}>$  
that may evolve upon the reception of an input on $x$ as follows:
\vspace{-0.3cm}
$$\infer [{[\inpconstset]}]{<\{\depen{x}{i}\};\{\constr{y}{c}{1}{\{\ind{x}\}},\constr{y'}{v}{1}{\emptyset}\}>\xrightarrow{\text{$\depen{x?}{i}$}}<\{\depen{x}{i}\};\{\constr{y}{c}{0}{\emptyset},\constr{y'}{v}{1}{\emptyset}\}>}{\infer[{[\inpinit]}]{\constr{y}{c}{1}{\{\ind{x}\}}\xrightarrow{\text{$x?$}}\constr{y}{c}{0}{\emptyset}}{} & \infer[{[\inpnotdom]}]{\constr{y'}{v}{1}{\emptyset}\xrightarrow{\text{$x?$}}\constr{y'}{v}{1}{\emptyset}}{x\notin dom[\D]}}$$
\end{example}


 
     

 




The type language serves as a means to capture component behaviour, and types for components may be obtained (inferred) as explained below. 
The results presented afterwards ensure that when the type extraction is possible, then each behaviour in the component is explained by a behaviour in the type, and 
that each behaviour in the type can eventually be exhibited by the component.

\subsection{ Type extraction for base components}\label{sec:base}

We describe the procedure that allows to (automatically) extract the type of a component, focusing first on the case of base components, remembering their reactive flavour.
The goal is to identify the basic types associated to the communication ports, as well as the dependencies between them, while checking that their usage is consistent throughout. 

In order to extract the type of a base component we need to define two auxiliary functions. First, we assume that from each function
$f(\tilde{x})$ used in a local binder, we can infer the respective function type. 
We introduce the notation $\gamma(\cdot)$ to represent a mapping from basic elements (such as values, ports, or functions) to their respective types.
We also use $\gamma$ for lists of elements in which case we obtain the list of respective types (e.g., $\gamma(1,\mathtt{hello})=\mathit{integer},\mathit{string}$). 
Second, given a local binder $y=f(\tilde{x})<\tilde{\sigma}$, we need to count the number of values that $y$ has available at runtime for each of the ports in $\tilde{x}$.
This corresponds to the number of elements in $\tilde{\sigma}$ that have a mapping for a port $x$ to a value, which we denote by $\tcount{x}{\tilde{\sigma}}$ defined as follows.
Let $X$ be the set of ports and $\Sigma$ a set whose elements are the lists of mappings from ports to values. Then function $count:X\times \Sigma\rightarrow \N_0$ is defined as follows:
  
 $\tcount{x}{\tilde{\sigma}}= $ 
 $\begin{cases}
 
 j & \mbox{if } \ \tilde{\sigma} = \sigma_1,\dots,\sigma_j,\sigma_{j+1}, \ldots,\sigma_l \wedge x \in \bigcap_{1\leq i\leq j}\dom(\sigma_i) \ \wedge \  x \notin \bigcup_{j+1\leq i \leq l}\dom(\sigma_i)\\
 
 0 & \mbox{otherwise}
 \end{cases}$  \\ 
 
 Notice that mappings in $\tilde{\sigma}$  are handled following a FIFO discipline, so the first (oldest) mappings are the ones that need to be accounted for.
We may now define our type extraction procedure for base components:

\begin{definition}[Type Extraction for a Base Component]\thlabel{extractbase}
\hfill

Let $[\tilde{x} > \tilde{y}]\{y_1 =f_{y_1}( \tilde{x}^{y_1})<\tilde{\sigma}^{y_1}, \dots , y_k = f_{y_k}( \tilde{x}^{y_k} ) < \tilde{\sigma}^{y_k} \}$ be a base component, where $\tilde{y}=y_1,y_2,\dots,y_k$.
If there exists $\gamma$ such that $\gamma(\tilde{x})=\tilde{b}$ and $\gamma(y_1) = b_1', \dots, \gamma(y_k) = b_k'$ and provided that $\gamma(f_{y_i}) = \tilde{b}^{y_i} \rightarrow b_i'$ for any $i \in 1, \dots , k$ and that $\tilde{b}^{y_i} = \gamma(\tilde{x}^{y_i})$ for any $i \in 1,\dots , k$ then the extracted type of the base component is
$< X_b ; \C >$ 
where $X_b=\{ \depen{x}{b} \  | \  x \in \tilde{x} \wedge b=\gamma(x)\}$
and $$\C = \{\depen{y_i}{b_i’}:\infty:\D_{y_i} \ |\  i \in 1, \dots , k \wedge b_i'=\gamma(y_i) \wedge \D_{y_i} = \{ x : \tcount{x}{\tilde{\sigma}^{y_i}} \ | \  x \in \tilde{x}^{y_i} \}\}$$
\end{definition}

In \thref{extractbase} the list of local binders is specified in such a way that each function ($f_{y_i}$), its parameters ($\tilde{x}^{y_i}$) and the list of mappings ($\tilde{\sigma}^{y_i}$) are indexed with the output port that is associated to them ($y_i$), so as to allow for a direct identification. Moreover, notice that each list of arguments $\tilde{x}^{y_i}$ (of function $f_{y_i}$) is a permutation of list $\tilde{x}$, as otherwise they would be undefined. 
Notice also that every output port of the interface of the component has a local binder associated to it and that there is no local binder $y_t = f_{y_t}( \tilde{x}^{y_t} )< \tilde{\sigma}^{y_t}$ such that $y_t$ is not part of the component interface, i.e., we do not type components that have undefined output ports or that declare unused local binders, respectively. 
We also rely in \thref{extractbase} on (the existence of) $\gamma$ to ensure consistency. Namely, we consider $\gamma$ provides the list of basic types for the input ports ($\gamma(\tilde{x})=\tilde{b}$) and for the output ports ($\gamma(y_1) = b_1', \dots , \gamma(y_k) = b_k'$). Then, we require that $\gamma(f_{y_i})$, for each $f_{y_i}$, specifies the function type where the return type matches the one identified for $y_i$ (i.e., $b_i'$). Furthermore, we require that the types of the parameters given in the function type ($\tilde{b}^{y_i}$) match the ones identified for the respective (permutation of) input port parameters ($\gamma(\tilde{x}^{y_i})$).

We then have that the extracted type of a base component is a composition of two elements. The first one ($X_b$) is a set of input ports which are associated with their basic types. The second one is a set of constraints $\C$, one for each output port and of the form $\constr{y_i}{b_i'}{\infty}{\D_{y_i}}$. The constraint specifies the basic type ($b_i'$) which is associated to the output port, and the maximum number of values that can be output on $y_i$ is unbounded ($\infty$), since local binders can potentially perform computations indefinitely.
The third element of the constraint ($\D_{y_i}$) is a set of per each value dependencies (of port $y_i$) on the input port parameters $\tilde{x}^{y_i}$, capturing that each value produced on $y_i$ depends on a value being  received on all of the ports in $\tilde{x}^{y_i}$. Notice that the number of values that $y_i$ has available (at runtime) for each $x$ in $\tilde{x}^{y_i}$ is given by $\tcount{x}{\sigma^{y_i}}$. 

From an operational perspective, \thref{extractbase} can be implemented by first considering the type inferred for the functions in the local binders and then propagating (while ensuring consistency of) this information. 

\begin{example}
Consider our running example from Section~\ref{example}, in particular, component $K_{Portal}$ specified as $\interface{x_p,x_p'}{y_p,y_p',y_p''}\{y_p=f_u(x_p)<\tilde{\sigma}^{y_p},y_p'=f_r(x_p')<\tilde{\sigma}^{y_p'},y_p''=f()<\cdot\}$.


Let us take $\gamma$ such that  $\gamma(x_p,x_p')=i,c$ and $\gamma(y_p)=i$, $\gamma(y_p')=c$ and $\gamma(y_p'')=v$.
We know that function $f_u$ takes an $\mathit{image}$ ($i$) and gives an $\mathit{image}$ in return, hence $\gamma(f_u)=i\rightarrow i$.
Similarly, we also know that function $f_r$ is typed as $\gamma(f_r)=c\rightarrow c$. Function $f$ does not have any parameters hence $\gamma(time) = () \rightarrow v$. The extracted set of input ports with their types is $X_b=\{\depen{x_p}{i},\depen{x_p'}{c}\}$. Assume that the component is in the initial (static) state, so the queues of lists of mappings are empty (i.e., $\tilde{\sigma}^{y_p} = \cdot = \tilde{\sigma}^{y_p'}$).
Hence, we have that $ \tcount{x_p}{\tilde{\sigma}^{y_p}}=0$ and
 $\tcount{x_p'}{\tilde{\sigma}^{y_p'}}=0$. The extracted set of constraints is then $\C=\{\constr{y_p}{i}{\infty}{\{\pev{x_p}{0}\}},\constr{y_p'}{c}{\infty}{\{\pev{x_p'}{0}\}},\constr{y_p''}{v}{\infty}{\emptyset}\}$ and the extracted type of the component $K_{Portal}$ is $<X_b;\C>$.

\end{example}


\subsection{ Type extraction for composite components}\label{sec:composite}

Extracting the type of a composite component is more challenging than for a base component. The focus of the extraction procedure is on the interfacing subcomponent, which interacts both via forwarders and via the protocol. 
For the purpose of characterising how components interact in a given protocol, we introduce local protocols $LP$ which result from the projection of a (global) protocol to a specific role that is associated to a component. We reuse the projection operation from~\cite{Hugo}, where message labels are mapped to communication ports (thanks to distribution binders $D$) and also to basic types that describe the communicated values (that can be inferred from the ones of the ports). The syntax of local protocols $LP$ is: 
\begin{center}
$LP:= \tlp{x?}{\bt}.LP
       \ |\ \tlp{y!}{\bt}.LP
       \ |\ \mu \recvar X.LP
       \ |\ \recvar X
       \ |\ \tend$. 
\end{center}       
       Term $\tlp{x?}{\bt}.LP$ denotes a reception of a value of a type $\bt$ on port $x$, upon which protocol $LP$ is activated. Term $\tlp{y!}{\bt}.LP$ describes an output in similar lines. Then we have standard constructs for recursion and for specifying termination ($\tend$). 
       Our local protocols differ from the ones used in~\cite{Hugo} since here we only consider choice-free global protocols. To simplify the setting, we consider global protocols that have at most one recursion (consequently also the projected local protocols). We also consider that message labels can appear at most once in a global protocol specification (up to unfolding of recursion), hence also ports occur only once in projected local protocols (also up to unfolding). 
       
       We omit the definition of projection and present the intuition via an example.
  
  \begin{example} Let $G$ be the (one-shot) protocol $G=\gcom {Portal} {image} {RE};\gcom {RE} {class} {Portal};\gend$ from  Section~\ref{example} and let $\gamma(image,class)=i,c$ be a function that given a list of a message labels returns a list of their types.
  
  Then, the projection of protocol $G$ to role $\role {Portal}$, denoted by $G\downarrow_{\role{Portal}}$ is  protocol $\tlp{y_p!}{i}.\tlp{x_p'?}{c}.\tend$ and the projection of $G$ to role $\role {RE}$ is local protocol $\tlp{x_{re}?}{i}.\tlp{y_{re}!}{c}.\tend$, where ports $x_p',y_p,x_{re},y_{re}$ are obtained via distribution binders $\dbinder{image}{RE}{x_{re}} {\roleport{Portal}{y_p}},\dbinder{class}{Portal}{x_p'} {\roleport{RE}{y_{re}}}$. 
  Essentially, the local protocol of $\role{Portal}$ describes that first it emits an image on $y_p$ and then receives a classification on $x_p'$, and the local protocol of $\role{RE}$ says that it first receives an image on $x_{re}$ and then outputs a result of a classification on $y_{re}$.
  \end{example}

We introduce some notation useful for the definition of the type extraction for composite components.
We use the language context for local protocols (excluding recursion), denoted by $\context$, so as to abstract from the entire local protocol and focus on specific parts and we define it as:
$\context[\  \cdot \ ]::=x?:\bt.\context[\ \cdot \ ]\ |\ y!:\bt.\context[\ \cdot \ ]\  |\ \cdot$.
%
%
We denote the set of ports appearing in a local protocol by $\fp(LP)$ and by $\rep(LP)$ the set of ports that occur in a recursion  (e.g. in $LP$ for recursion $\mu \recvar X.LP$). Considering a list of forwarders $F$, we define two sets: by $\FX$ we denote the set of (internal) input ports and by $\FY$ the set of (internal) output ports which are specified in $F$ (e.g., if $F=\fbinder {x_p}x\ $ then $\FX = \{x_p\}$). 

We now introduce the important notions that are used in our type extraction, namely that account for \emph{values flowing} in a protocol and for the \emph{kinds of dependencies} involved in composite components. 
Finally, we address the \emph{boundaries} for the output ports.
\paragraph{Values flowing} Our types track the dependencies between output and input ports, including per each value dependencies that specify how many values received on the input port are available to the output port. As discussed in the previous section, for base components this counter is given by the number of values available in the local binder queues. For composite components, as preliminary discussed in Section~\ref{example}, per each value dependencies might actually result from a chain of dependencies that involve subcomponents and the protocol. So, in order to count how many values are available in such case, we need to take into account how many values are in the subcomponents (which is captured by their types) and also if a value is \emph{flowing} in the protocol. We can capture the fact that a value is flowing by inspecting the structure of the protocol. In particular we are interested in values that flow from $y$ to $x$ when an output on $y$ precedes an input in $x$ in a recursive protocol, hence when the protocol is of the form $\context[\mu \recvar X.\context'[\tlp{y!}{\bt'}.\context''[\tlp{x?}{\bt}.LP']]$. The value is flowing when the output has been carried out but the input is yet to occur, which we may conclude if the protocol is \emph{also} of the form $\context'''[\tlp{x?}{\bt}.LP'']$ where $x,y\notin \fp(\context'''[\ \cdot\ ])$. 
We denote by $\vf(LP,x,y)$ that there is a value flowing from $y$ to $x$ in $LP$, in which case
$\vf(LP,x,y)=1$, otherwise $\vf(LP,x,y)=0$. We will return to this notion in the context of the extraction of the dependencies of the output ports, discussed next.  



\begin{figure}[h!]
\centering
\begin{minipage}{.5\textwidth}
  \centering
  \includegraphics[width=0.5\linewidth]{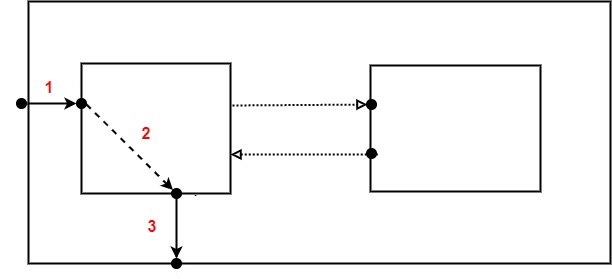}
  \caption{Direct Dependency}
  \label{fig:direct}
\end{minipage}%
\begin{minipage}{.5\textwidth}
  \centering
  \includegraphics[width=0.5\linewidth]{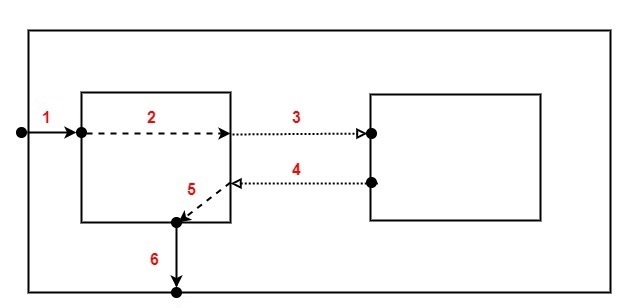}
  \caption{Transitive Dependency}
  \label{fig:transitive}
\end{minipage}
\end{figure}

\paragraph{Kinds of dependencies}
Composite components comprise two kinds of dependencies between output ports and input ports, illustrated in Figure~\ref{fig:direct} and Figure~\ref{fig:transitive}, which are dubbed direct and transitive, respectively.

We gather the set of direct dependencies, i.e., when external output ports directly depend on external input ports (see Figure~\ref{fig:direct}), 
in $\dd$ which is defined as follows:
 \[ \dd\triangleq\{\pev{x}{M}\ |\ \C=\{\constr{y}{\bt^y}{\B}{\{\pev{x}{M}\}\uplus\D}\}\uplus\C' \wedge x \in \FX  \wedge y \in \FY \}\] 
Hence, in $\dd$ we collect all the dependencies for $y$ given in (internal) constraint $\C$ whenever both ports are external and preserving the kind of dependency $M$ so as to lift it to the outer interface.
 
For transitive dependencies (see Figure~\ref{fig:transitive}) to exist there are three necessary conditions. The first condition is to have in the description of a local protocol at least one output action, say on port $y'$, that precedes at least one input action, say on port $x'$. The second condition is that such output port $y'$ depends on some external input port $x$ and the third condition is that there exists some external output port $y$ that depends on the input on $x'$. In such cases, we say that $y$ depends on $x$ in a transitive way. 
 
 
 We introduce a relation that allows to capture the first condition above.
Let $LP$ be the local protocol that is prescribed for an interfacing component. Two ports $x'$ and $y'$ are in relation $\rel{LP}{i}$ for some local protocol $LP$
if $x',y'\in \fp(LP)$ and where $i\in \{1,2,3\}$ as follows: $y'\rel{LP}{1} x'$  if  $LP=\context[\tlp{y'!}{\bt^{y'}}.\context'[\tlp{x'?}{\bt^{x'}}.LP']]$ and $x',y'\notin \rep(LP)$;
$y'\rel{LP}{2} x'$  if  $LP=\context[\tlp{y'!}{\bt^{y'}}.\context'[\mu \recvar X.\context''[\tlp{x'?}{\bt^{x'}}.LP']]]$ and $y'\notin \rep(LP)$;
$y'\rel{LP}{3} x'$  if  $LP=\context[\mu \recvar X.\context'[\tlp{y'!}{\bt^{y'}}.\context''[\tlp{x'?}{\bt^{x'}}.LP']]]$.
We distinguish three cases: when both the output and the input are non-repetitive, when only the input is repetitive, and when both the output and the input are repetitive. 

We may now characterise the transitive dependencies.  
Let $\chorbox {\tilde x'}{\tilde y'}{D}{\role{r}[F]}{r=K,R}{G}$ be the composite component, $T_r=<X_b,\C>$ the type of interfacing component $K$ and $LP$ its local protocol. 
The set of transitive dependencies on $y$, denoted $\dt$, is defined relying on an abbreviation $\eta$ as follows:

\begin{tabular}{l l l}
 $\eta$ & $=$& $\C=\{\constr{y}{\bt_1}{\B}{\{\pev{x'}{M'}\}\uplus\D'},\constr{y'}{\bt_2}{\B'}{\{\pev{x}{M}\}\uplus\D}\}\uplus\C'$\\
& & $  \wedge\  x\in \FX \wedge y\in \FY \wedge y'\rel{LP}{i} x'$ 
 \\\\
 $\dt$ &$ \triangleq$ & $\big\{\ind{x}  \ \ |\ \ \eta \wedge i\in\{1,2\}\wedge M\ngtr0  \} $\\
  & & $\bigcup $\\
& & $\big\{\ind{x}  \ \ |\ \ \eta \wedge i=3 \wedge ( M=\ic \lor ( M'=\ic \wedge M=0 \wedge \vf(LP,x',y')=0)) \big\}$\\
 & & $\bigcup $\\
 
 

& & $\big\{\pev{x}{(N+N'+\vf(LP,x',y'))}\ \ |\ \ \eta \wedge i=3 \wedge M=N \wedge M'=N' \big\}$

\end{tabular}

In $\eta$ we gather a conjunction of conditions that must always hold in order for a transitive dependency to exist: namely that the (internal) constraint $\C$ specifies dependencies between 
$y$ and $x'$ and between $y'$ and $x$ and also that $y$ and $x$ are external ports while $y'$ precedes $x'$ in the protocol. To simplify presentation of the definition of $\dt$ we rely on 
the (direct) implicit matching in $\eta$ of the several mentioned elements.

There are two kinds of transitive dependencies that are gathered in $\dt$, namely initial ($\ind{x}$) and per each value ($\pev{x}{N}$). For initial dependencies there are two separate cases to consider. 
The first case is when the protocol specifies that the output on $y'$ is non-repetitive ($i \in \{1,2\}$), hence will be provided only once. Condition $M\ngtr0$ says that no values 
are already available for that initial output to take place (internally to the component that provides them as specified in $\eta$), hence either $M = \ind{x}$ or $M = 0$. 

The second case for an initial transitive dependency is when both $y'$ and $x'$ are repetitive in the protocol ($i = 3$) but at least one of the internal dependencies (between $y'$ and $x$ and between $y$ and $x'$, given by $M$ and $M'$ respectively) is an initial dependency. This means that, regardless of the protocol, such a dependency is dropped as soon as a value is provided which implies that the transitive dependency is also dropped. Since $M$ is at the beginning of the dependency chain, if it is initial then no further conditions are necessary. However, if $M'$ is initial we need to ensure that there is no value already flowing ($\vf(LP,x',y')=0$) or already available to be output on $y'$ ($M = 0$), since only in such case (an initial) value is required from the external context (i.e., otherwise if $\vf(LP,x',y')=1$ or $M \geq 0$ then the chain of dependencies is already ``internally'' satisfied).


Finally, we have the case of per each value transitive dependency, that can only be when both $y'$ and $x'$ are repetitive in the protocol ($i = 3$) and internal dependencies $M$ and $M'$ are both per each value dependencies ($M = N$ and $M'=N'$), which means that the dependency chain is persistent. The number of values available of (external) $x$ for $y$ is the sum of the values available in the internal dependencies ($N$ and $N'$) plus one if there is a value flowing (zero otherwise).
%
%
 %
Notice that the definition of value flowing presented previously focuses exclusively in the case when $y'$ and $x'$ are repetitive in the protocol, since this is the only case where values might be flowing and the dependency is still present in the protocol structure (i.e., $y'\rel{LP}{3} x'$ holds). In contrast, a dependency $y'\rel{LP}{i} x'$ for $i\in \{1,2\}$ is no longer (structurally) present as soon as the value is flowing (i.e., a non-repetitive $y'$ no longer occurs in the protocol after an output).

It might be the case that one output port depends in multiple ways on the same input port. For that reason we introduce a notion of \textit{priority} among dependencies, denoted by $\pr(\ ,\ )$ 
that gives priority to per each value dependencies (with respect to ``initial''). 
%
The priority builds on the property that if multiple per each value dependencies (including direct and transitive) are collected (e.g., $x:N_1, \ldots, x:N_k$) then the number of available values specified in them is the same (i.e., $N_1 = \ldots = N_k$).
The list of dependencies for  port $y$ is then given by the  (prioritised) union of direct and transitive dependencies: \vspace{-0.3cm} \[\D(\C,F,LP,y)=\pr(\dd \cup \dt)\] \vspace{-1.2cm}

\paragraph{Boundaries}
The last element that we need to determine in order to extract the type of a composite component is the boundary of output ports. 
The type of the interfacing component already specifies a (internal) boundary, however this value may be further bound by the way in which the component is used in the composition.
In particular, if an output port depends on input ports that are not used in the protocol nor are linked to external ports, then no (further) values are received in them and the potential for the output port is consequently limited.
We distinguish three cases for three possible limitations:

$B_1=\{N' \ | \  \C=\constr{y}{\bt_1}{\B}{\{\pev{x'}{N'}\}\uplus\D'}\uplus\C'\wedge x'\notin(\fp(LP)\cup \FX)\}$  

$B_2=\{0 \ | \  \C=\{\constr{y}{\bt_1}{\B}{\{\ind{x'}\}\uplus\D'}\}\uplus\C'\wedge x'\notin(\fp(LP)\cup \FX)\}$ 

$B_3=\{ (N'+1) \ | \ \C=\{\constr{y}{\bt_1}{\B}{\{\pev{x'}{N'}\}\uplus\D'}\}\uplus\C'\wedge x'\in \fp(LP) \wedge x'\notin(\rep(LP) \cup \FX)\}$ 

In $B_1$ and $B_2$ we capture the case when there is a dependency on a  port that is not used in the protocol ($x'\notin \fp(LP)$) nor linked externally ($x'\notin \FX$), where the difference is in the kind of dependency. For per each value dependencies (if any), the minimum of the internally available values is identified as the potential boundary, while for initial dependencies (if present) the potential boundary is zero (or the empty set).
In $B_3$ we capture a case similar to $B_1$ where the port is used in the protocol but in a non-repetitive way, hence only one (further) value can be provided.
%

The final boundary determined for $y$, denoted by $\B(y,LP,\C)$, is the minimum number among the internal boundary of $y$ (i.e., $\B$ if $\C = \constr{y}{\bt}{\B}{\D}  \uplus \C'$) and possible boundaries $B_1,B_2$ and $B_3$ described above.
 \vspace{-0.3cm}\[\B(y,LP,\C)=min(\{\B\} \cup B_1 \cup B_2 \cup B_3)\] \vspace{-0.8cm} 

We may now present the definition of type extraction of a composite component relying on a renaming operation. 
Since the type extraction of a composite component focuses on the interfacing subcomponent, we single out the ports that are linked via forwarders to the external environment. To capture such links, we introduce renaming operation $\ren(\ ,\ )$ that renames the ports of the interfacing subcomponent to the outer ones by using the forwarders as a guideline.
For example, if we have that $F=\fbinder {x_p}x \ $ than $\ren(F,x_p)=x$. 

\begin{definition}[Type Extraction for a Composite Component]
\thlabel{compositeextraction} 

Let $\chorbox {\tilde x}{\tilde y}{D}{\role{r}[F]}{r=K,R}{G}$ be a composite component and $LP=G\downharpoonright_r$ the local protocol for component $K$. If $T_r=<X^r_b;\C^r>$ is the type of component $K$, then the extracted type from $LP$ and $T_r$ is \vspace{-0.3cm}\[T(LP,T_r,F)=\ren(F,<X_b;\C>)\] where: 
$X_b=\{\depen{x}{\bt}\ |\ 
\depen{x}{\bt} \in X^r_b 
\wedge x\in \FX\}$ 

$\ \ \ \ \ \ \ \C=\{\constr{y}{\bt'}{\B(y,LP,\C^r)}{\D(\C^r,F,LP,y)}\ |\ 
\C^r = \{\constr{y}{\bt'}{\B'}{\D'}\}  \uplus \C' 
\wedge y\in \FY \}$. 
\end{definition}

\begin{example}
Let us extract the type of component $K_{IRS}$ from Section~\ref{example} considering protocol  $G=\gcom {Portal} {image} {RE};\gcom {RE} {class} {Portal};\gend$. The type of interfacing component $K_{Portal}$ is \[T_{Portal}=<\{\depen{x_p}{i},\depen{x_p'}{c}\};\{\constr{y_p}{i}{\infty}{\{\pev{x_p}{N_p}\}},\constr{y_p'}{c}{\infty}{\{\pev{x_p'}{N_p'}\}},\constr{y_p''}{v}{\infty}{\emptyset}\}>\] 

We have that local protocol is $LP=\tlp{y_p!}{i}.\tlp{x_p'?}{c}.\gend$ and sets of external ports  $\FX=\{x_p\}$ and $\FY=\{y_p',y_p''\}$, where $\ren(F,x_p)=x$, $\ren(F,y_p')=y$ and $\ren(F,y_p'')=y'$.
This immediately gives us the set of input ports that is in the description of the type of component $K_{IRS}$ which is $\ X_b=\{\depen{x}{i}\}$.

Let us now determine the constraints of the output ports. 
Since port $y_p''$ has no dependencies also port $y'$ will not have any, and moreover has the same boundary ($\infty$). So, the extracted constraint for $y'$ will be $\ren(\constr{y_p''}{v}{\infty}{\emptyset})$, which is $\constr{y'}{v}{\infty}{\emptyset}$.
Port $y_p'$ instead depends on port $x_p'$ which is used in the protocol ($x_p' \in\fp(LP)$). Since the protocol is not recursive we have the consequent limited boundary (case $B_3$ explained above), namely
 the boundary of $y_p'$ is $min(N_p' + 1,\infty)=N_p'+1$. 
Furthermore. we have that $y_p\rel{LP}{1}x_p'$ and that $y_p$ has per each value dependency $\pev{x_p}{N_p}$. If $N_p>0$ then $y_p'$ does not transitively depend on $x_p$, otherwise there is an initial dependency. 
Let us consider the initial (static) state where no image has been receive yet, i.e., $N_p=0$. In such case we have that the resulting constraint for $y_p'$ is $\constr{y_p'}{c}{N_p'}{\ind{x_p}}$, which after renaming for $y$ 
is $\constr{y}{c}{N_p'}{\ind{x_p}}$. So, the extracted type of $K_{IRS}$ is the following \[\{\depen{x}{i}\};\{\constr{y}{c}{N_p'}{\ind{x_p}},\constr{y'}{v}{\infty}{\emptyset}\}\]

\end{example}


\subsection{Type Safety}\label{safety}

In this section we present our main results that show a tight correspondence between the behaviour of components and of their extracted types.
Apart from the conditions already involved in the type extraction, for a component to be well-typed we must also ensure that any component that interacts in a protocol can actually
carry out the communication actions prescribed by the protocol. 

 
 For this reason we introduce the conformance relation, denoted by $\bowtie$, that asserts compatibility 
 between the type of a component and the local protocol that describes the communication actions prescribed for the component.
For the purpose of ensuring compatibility, in particular for the interfacing component, we also introduce an extension of our type language,
dubbed modified types $\itype$ (see Appendix~\ref{apdxmod}). The idea for modified types is to allow to abstract from input dependencies 
from the external environment, namely by considering such dependencies can (always) potentially be fulfilled, allowing conformance to focus on internal compatibility. 
By $\itype(F,T)$ we denote the modified type that results from abstracting such external dependencies in $T$, relying on forwarders $F$, namely by considering per each value dependencies 
for external input ports are unbounded ($\infty$) and dropping initial dependencies.
The rules for the semantics of modified types are the same as the ones shown in Section~\ref{language}, the only implicit difference for modified types is that decrementing an unbounded dependency has no effect.

The definition of the conformance relation (see Appendix~\ref{apendixconformance}) is given by induction on the structure of the local protocol and it is characterised by judgments of the form $\Gamma \vdash \itype \bowtie LP$, where $\Gamma$ is a type environment that handles protocol recursion (i.e., $\Gamma$ maps recursion variables to modified types). 
We report and comment here only on the rules for input and output:  
\[
\infer[{[InpConf]}]{\Gamma \vdash \itype \bowtie \tlp{x?}{\bt}.LP}{\itype \inx \itype' & \Gamma \vdash \itype' \bowtie LP} \ \ \ \ \ \ \ \ \ \ 
\infer[{[OutConf]}]{\Gamma \vdash \itype \bowtie \tlp{y!}{\bt}.LP}{\itype \outy \itype' & \Gamma \vdash \itype' \bowtie LP}
\]
Rule [$InpConf$] states that a modified type $\itype$ is conformant with protocol $\tlp{x?}{\bt}.LP$, 
if $\itype$ can input a value of type $b$ on port $x$ and if continuation $\itype'$ is conformant 
with the continuation of protocol $LP$.
Rule [$OutConf$] is similar but deals with the output of a value on port $y$. 

We can now formally define when a component $K$ has type $T$, in which case we say $K$ is well-typed.
\begin{definition} Let $K$ be a component, we say that $K$ has a type $T$, denoted by $K\Downarrow T$:
\begin{enumerate}
   \item If $K$ is a base component, $K\Downarrow T$ when $T$ is obtained by \thref{extractbase}.
        
   \item If $K=[\tilde{x}>\tilde{y}]\{G;\role {r_1}=K_1, \dots, \role {r_k}=K_k;D;\role {r_1}[F]\}$ then 
         $K\Downarrow T$ when

   \begin{itemize}
     \item $\exists T_{r_i}\ |\ K_i\Downarrow T_{r_i}, \text{for } i=1,2,\dots,k$; 
            
     \item $T$ is extracted type from $T_{1}$ and $G\downharpoonright_{r_1}$ by \thref{compositeextraction}; 
            
     
     \item $\itype(F,T_{r_i})\bowtie G\downharpoonright_{r_i} \text{for } i=1,2,\dots,k$;    
   \end{itemize}
    \end{enumerate}
\end{definition}

Notice that the definition relies on modified types for conformance, but for any type $T$ not associated with the interfacing component we have 
that $\itype(F,T) = T$ since there can be no links to external ports (assuming that all ports have different identifiers). 

%

We can now our type safety results given in terms of Subject Reduction and Progress,
which provide the correspondence between the behaviours of well-typed components and their types.
In the statements we rely on notation $\lambda(v)$ that represents $x?(v)$, $y!(v)$ or $\tau$ 
and $\lambda(\bt)$ that represents $x?(\bt)$, $y!(\bt)$ or $\tau$.

\begin{theorem}[Subject Reduction]\thlabel{SubjectReduction}
If
    $K \Downarrow T$ and $K\xrightarrow{\textit{$\lambda$(v)}}K' $ and $v$ has type $\bt$ 
    then
     $T\xrightarrow{\textit{$\depen{\lambda}{\bt}$}}T'$ and $K'\Downarrow T' $.    
\end{theorem}
\vspace{-0.5cm}
\begin{proof} 
By induction on the derivation of $K\xrightarrow{\text{$\lambda (v)$}} K'$.
\end{proof}

\thref{SubjectReduction} says that if a well-typed component $K$ performs a computation step to $K'$, 
then its type $T$ can also evolve to type $T'$ which is the type of component $K'$. 
Moreover, the theorem ensures that if $K$ carries out an input or an output of a value $v$, type $T$ performs 
the corresponding action at the level of types.
\thref{SubjectReduction} thus attests that well-typed components always evolve to well-typed components, and furthermore that any component evolution can be described by an evolution in the types.

The progress result does not describe a strong correspondence like for Subject Reduction since we need to abstract from internal computations in components.
For that reason, in the Progress statement we rely on $K\xRightarrow{\textit{$\lambda$(v)}}K'$ to denote a sequence of transitions  
$K \xrightarrow{\tau} {\cdots}K''\xrightarrow{\lambda(v)}K'''\xrightarrow{\tau}{\cdots} K'$, i.e, that component 
$K$ may perform a sequence of internal moves, then an I/O action, after which another sequence of internal moves leading to $K'$.
\begin{theorem}[Progress]\thlabel{Progress}
If
    $K \Downarrow T$ and $T\xrightarrow{\textit{$\depen{\lambda}{\bt}$}}T'$  and $\lambda(\bt) \neq \tau$ 
    then  $\bt$ is the type of a value $v$ and
    $K\xRightarrow{\textit{$\lambda$(v)}}K'$ and $K'\Downarrow T'$.
\end{theorem}
\vspace{-0.5cm}
\begin{proof}
By induction on the structure of $K$.
\end{proof}

\thref{Progress} says that if type $T$ of component $K$ can evolve by exhibiting an I/O action to type $T'$, then $K$ can eventually (up to carrying out some internal computations) 
exhibit a corresponding action leading to $K'$, and where $K'$ has type $T'$. 
\thref{Progress} thus ensures that the behaviours of types can eventually be carried out by the respective components,
which entails components are not stuck and allows, together with \thref{SubjectReduction}, to attest that types faithfully capture component behaviour.
Intuitively, our types can be seen as \emph{promises of behaviour} in the sense that whatever they prescribe as possible behaviours, the components will eventually deliver. 
For the sake of addressing any possible component configuration, in particular when components have already all the dependencies (internally) satisfied in order to provide some behaviour,
it is crucial that types capture the number of (internally) available resources.

\section{Concluding Remarks}\label{finish}

In this paper we introduce a type language for the choice-free subset of the GC language~\cite{Hugo} that characterises the reactive behaviour of components and allows to capture ``what components can do". 
In particular, our types describe the ability of components to receive and send values, while 
tracking different kinds of dependencies (per each value and initial ones) and specifying constraints on the boundary of the number of values that a component can emit.
We show how types of components can be extracted (inferred) and prove that types faithfully capture component behaviour by means of Subject Reduction and Progress theorems.
Typing descriptions such as ours are crucial to promote component reusability, since to use a component we should only need to analyse its type and not its implementation (like in~\cite{Hugo}).
For instance, for the sake of ensuring the behaviour of a component is compatible with a governing communication protocol, where such compatibility is attested in our case by the conformance 
of the type to the (local) protocol.

We place our approach in the behavioural types setting (cf.~\cite{DBLP:journals/csur/HuttelLVCCDMPRT16}) since our types evolve in order to explain component behaviour (cf. \thref{SubjectReduction}), in contrast with classic subject reduction results where the type is preserved.
In the realm of behavioural types, we distinguish Multiparty Asynchronous Session Types~\cite{DBLP:journals/jacm/HondaYC16} which actually lay the basis for the protocol language of our target model~\cite{Hugo}.
The model builds on the idea that protocols can be used to directly program the interaction, and not only serve as a specification/verification mechanism, following the approach of 
choreographic programming~\cite{DBLP:conf/popl/CarboneM13, fabrizio}.

We discuss some closely related work, starting by Open Multiparty Sessions~\cite{OMS} which to some extent shares the same goals and the same background (cf.~\cite{DBLP:journals/jacm/HondaYC16}).
The approach in~\cite{OMS} targets the composition of protocols by considering that one of the participants can actually be instantiated by an external environment. Two protocols can then be connected 
if there is a participant in each that can serve as the interface to the other interaction. So protocols can be viewed as the units of composition instead of components like in our case, and reusing such protocols
in other compositions requires compatibility between the I/O actions which are prescribed for the interfacing role. The main difference is therefore that we consider components that are potentially more reusable
considering the I/O flexibility provided the reactive flavour.



We also identify the CHOReVOLUTION~\cite{Chor}
 project where the assembly of services via a choreography is addressed. The I/O flexibility is provided by adapters at assembly time that can solve 
I/O interface mismatches between service and choreography. 
We remark that the CHOReVOLUTION approach is at a very mature state (including  tool support~\cite{chorevolution}), where however an assembly of services cannot be provided as a unit of reuse (like our composite components). Differently, our type-based approach aims at abstracting from the implementation and provides more general support for component substitution and reuse. Similar to our work, the model of \textit{service based architectures} (SOA's)~\cite{SOA} exposes components that exchange services between each other via ports. However, the authors do not present the type language, where some ideas we presented for extracting a type of a component could be used for the model they present.


We may also find the notion of distributed components is the Signal Calculus (SC)~\cite{SC}, where in each component a process is located. 
The type-based approach presented in~\cite{CEN} addresses the issue of ensuring SC local processes are composed and interact in a way such that they follow a well-defined protocol of interaction.
Our model embeds the protocol in the operational semantics so such coordination is ensured by construction. Our types focus on a different purpose of ensuring data dependencies are met in order to ensure components are not stuck in the sense of the progress result.

In the work about Interactive verification of ADPs~ \cite{Marmsoler2019}  the authors introduce a notion of component type which characterizes components with a certain behaviour. One core difference between \cite{Marmsoler2019} and our work is that the authors do not provide any means to automatically extract types from given components.   


We believe the ideas reported in this paper can contribute to the theoretical basis for providing support for component-based development in distributed systems. 
Immediate directions for future work include the support for protocols with branching, and providing a characterisation of the substitution principle~\cite{DBLP:journals/toplas/LiskovW94} based in our types.
Further challenges remain at the level of conveying the theoretical model to concrete applications, in particular regarding component deployment and the support for their persistent reuse.
\vspace{-0.4cm}
 \paragraph{Acknolegements} We thank the anonymous reviewers  for their suggestions and comments which helped us to improve the paper.

\bibliographystyle{eptcs}

 \bibliography{bib}

\appendix

\section{Modified type $\itype$}\label{apdxmod}

Now we introduce the modified type denoted by $\itype$. The interfacing component of the composite one, beside its interaction with other components, also interacts with an external environment. In this case the crucial part is that it is able to receive in any moment values that are input externally. For the purpose of observing if a type of a component can perform actions required by the protocol, we need to modify the type according to the possible inputs that a (interfacing)  component can receive from the external context without any constraints. The modified type of a type $T$, taking into account the list of corresponding list of forwarders, is denoted by $\itype(F,T)$. If $T$ is the type of the interfacing component, each dependency on the external input ports is per each value dependency and the number of values available is unbounded (assuming that whenever the value is available it is received on the external input ports). The syntax of $\itype$-type is given in the Table \ref{tab:modsyntax}. It is similar to a syntax of the types which we have already shown, with the difference in the number of values received, that in the modified type can be unbounded (infinite).  Moreover, the rules defining the semantics of modified type are the same as the ones shown for our typing language (Table~\ref{tab:modsemantics}).\\

\textbf{$\itype$-Type syntax}
\vspace{-0.5cm}

\begin{table}[H]
\centering
\small{
\begin{tabular}{c |c  |c }

Types  & Constraints  & Dependencies \\ $\itype\overset{\Delta}{=}\types{X_b}{\iC} $\\

$X_b\overset{\Delta}{=}\{\depen{x_1}{b_1}, \dots , \depen{x_k}{b_k}\} $

& $\iC\overset{\Delta}{=}\{\constr{y_1}{\bt_1}{\B_1}{\iD_1}, \dots , \constr{y_k}{\bt_k}{\B_k}{\iD_k}\}$  

&$\iD\overset{\Delta}{=}\{\pev{x_1}{\M_1}, \dots , \pev{x_k}{\M_k}\}$ \vspace{1mm}\\
\hline

 Kinds of Dependencies & Boundaries & \\
 $\M::=\iN\ |\ \ic$  & $\B::=N\ |\ \infty$ & $k\geq 0; N\in \mathbb{N}_0$\\
 $\iN::= N \ |\  \infty$ & &\\

\end{tabular}}

\caption {$\itype$-Type syntax} \label{tab:modsyntax}
\end{table}

\textbf{$\itype$-Type semantics}
\vspace{0.5cm}

\begin{table}[h!]
\begin{center}

 \small

$\begin{tabular}{c c}

  \infer[{[\iinpnotdom]}]{\constr{y}{\bt}{\B}{\iD}\xrightarrow{\text{$x?$}}\constr{y}{\bt}{\B}{\iD}}{x\notin dom[\iD]} &  \infer[{[\iinpinit]}]{\constr{y}{\bt'}{\B}{\{\ind{x}\}\uplus\iD}\xrightarrow{\text{$x?$}}\constr{y}{\bt'}{\B}{\iD}}{}\\

\infer[[{\iinppev]}]{\constr{y}{\bt'}{\B}{\{\pev{x}{\iN}\}\uplus\iD}\xrightarrow{\text{$x?$}}\constr{y}{\bt'}{\B}{\{\pev{x}{\iN+1}\}\uplus\iD}}{} & \infer[{[\iinternal]}]{\itype\xrightarrow{\text{$\tau$}}\itype}{}\\

  \multicolumn{2}{c}{\infer [{[\iinpconstset]}]{<\{\depen{x}{\bt^x}\uplus X_b\};\{\constr{y_i}{b_i}{\B_i}{\iD_i} | i \in 1,\dots ,k \}>\xrightarrow{\text{$\depen{x?}{\bt^x}$}}<\{\depen{x}{\bt^x}\uplus X_b\};\{\constr{y_i}{b_i}{\B_i}{\iD'_i} | i \in 1,\dots ,k \}>}{\forall i\in 1,2,\dots, k &  \constr{y_i}{b_i}{\B_i}{\iD_i}\xrightarrow{\text{x?}}\constr{y_i}{b_i}{\B_i}{\iD_i'}}}\\
 \multicolumn{2}{c}{ \infer [{[\ioutdepset]}]{<X_b;\{\constr{y}{\bt^y}{\B}{ \{\pev{x_i}{\iN_i}  | i \in 1,\dots ,k \}}\}\uplus \iC>\xrightarrow{\text{$\depen{y!}{\bt^y}$}}<X_b;\{\constr{y}{\bt^y}{\B-1}{\{\pev{x_i}{\iN_i-1}  | i \in 1, \dots ,k \}}\}\uplus \iC>}{\B>0 & \iN_i\geq 1}}  \\

 \end{tabular}$

\end{center}
 \vspace{-2mm}
\caption {$\itype$ Semantics}
\label{tab:modsemantics}
\end{table}

\vspace{-0.5cm}

\begin{definition}\thlabel{defmod}
If $T_r=<X_b;\C>$ is a type of interfacing subcomponent $\overline{K}$ of composite component $[\tilde{x}>\tilde{y}]\{G;r=\overline{K},R;D;r[F]\}$ then $\itype(F,T_r)$ is the $T_r$-modified type where:\\

\begin{tabular}{l l l l}

    \ $\itype(F,<X_b,\C>)$ & $\triangleq$ & $<X_b;\itype(F,\C)>$ &\\

      $\itype(F,\{\constr{y}{\bt}{\B}{\D}\}\uplus\C)$ & $\triangleq$ & $\itype(F,\{\constr{y}{\bt}{\B}{\D}\})\uplus\itype(F,\C)$ &\\

     $\itype(F,\{\constr{y}{\bt}{\B}{\D}\})$ & $\triangleq$ & $\{\constr{y}{\bt}{\B}{\itype(\D)}\}$ &\\

  $\itype(F,\{\pev{x}{M}\}\uplus \D)$ & $\triangleq$ & $\itype(F,\{\pev{x}{M}\})\uplus\itype(\D)$ & where $M\in\{N,\ic\}$
    \\
    
   $\itype(F,\pev{x}{M})$ & $\triangleq$ & $\{\pev{x}{M}\}$ & if $x\notin \FX$, where $M\in\{N,\ic\}$\\
   
    $\itype(F,\pev{x}{M})$ & $\triangleq$ & $\{\pev{x}{\infty}\}$ & if $x\in \FX$\\
     $\itype(F,\ind{x})$ & $\triangleq$ & $\emptyset$ & if $x\in \FX$

    
 

\end{tabular}
\end{definition}

Note that for $K=[\tilde{x}>\tilde{y}]\{G,r_1=K_1,r_2=K_2,\dots,r_n=K_n;D;r_1[F]\}$ we have that $$\itype(F,T_{r_2})=T_{r_2},\dots, \itype(F,T_{r_k})=T_{r_k}$$ since the only component that forwards the values from/to external environment is component $K_1$.

\section{Conformance relation}\label{apendixconformance}

\begin{table}[H]
    \centering
    \begin{tabular}{c c c}
    \multicolumn{3}{c}{   \infer[{[InpConf]}]{\Gamma \vdash \itype \bowtie \tlp{x?}{\bt}.LP}{\itype \inx \itype'  \Gamma \vdash \itype' \bowtie LP}  \ \ \ \  \infer[{[OutConf]}]{\Gamma \vdash \itype \bowtie \tlp{y!}{\bt}.LP}{\itype \outy \itype' & \Gamma \vdash \itype' \bowtie LP}}\\
       
       & & \\
       
        \infer[{[EndConf]}]{\Gamma \vdash \itype \bowtie \gend}{}  &  \infer[{[VarConf]}]{\Gamma,X:\itype' \vdash \itype \bowtie X}{\itype'\leq \itype} &
       
    \infer[{[RecConf]}]{\Gamma \vdash \itype \bowtie recX.LP}{\Gamma,X:\itype \vdash \itype \bowtie LP} \\ 
     & & \\
         
    \end{tabular}
    \vspace{-0.2cm}
    \caption{Conformance}
    \label{tab:conformance}
\end{table}

\begin{definition}\thlabel{modevolve}  $\itype' \leq \itype $  if exists a (possibly empty) set of typed input ports $\{\depen{x_1}{\bt_1},\depen{x_2}{\bt_2}, \dots, \depen{x_k}{\bt_k}\}$ such that $\itype'\xrightarrow{\text{$\depen{x_1?}{\bt_1}$}}\cdots\xrightarrow{\text{$\depen{x_k?}{\bt_k}$}}\itype$.

\end{definition}

\vspace{-0.1cm}

Rule [$InpConf$] ensures that a modified type $\itype$ is conformant with the protocol, where it can receive an input of a matching type with a continuation as a protocol $LP$, if a modified type can receive a value on port $x$, and assuming that port $x$ receives a values of type $\bt$ and the evolved type is conformant with $LP$. Similar reasoning is for an output. Rule [$EndConf$] states that a modified type is always conformant with the termination protocol. Finally, we have two rules [$VarConf$] and [$RecConf$] for the recursion. The premise of Rule~[$VarConf$] requires that the type associated with the recursion variable by assumption and the type under consideration are related as $\itype'\leq\itype$ (\thref{modevolve}). Observing the semantics of modified types, the possible difference between types $\itype'$ and $\itype$ is that some initial dependencies might be dropped or that the number of values available on some input ports for some outputs might increase. Rule~[$RecConf$] states that $\itype$ is conformant with a protocol $rec \recvar X.LP$, provided that the type is conformant with the body of the recursion under the environment extended with assumption $\recvar X: \itype$.

\end{document}